\documentclass[twocolumn]{aastex62}
\usepackage{natbib}

\graphicspath{{./}{figures/}}
\usepackage{mathtools}

\received{}
\revised{}
\accepted{}

\submitjournal{ApJ}

\begin{document}

\title{S62 on a 9.9-year orbit around SgrA*}

\correspondingauthor{Florian Pei{\ss}ker}
\email{peissker@ph1.uni-koeln.de}
%\uccode `ß="1E9E
%\author[0000-0002-0786-7307]{Florian Pei{\ss}ker}
%\author[0000-0002-9850-2708]{Florian Pei{"1E9E}ker}
\author[0000-0002-9850-2708]{Florian Pei$\beta$ker}
\affil{I.Physikalisches Institut der Universit\"at zu K\"oln, Z\"ulpicher Str. 77, 50937 K\"oln, Germany}
\author{Andreas Eckart}
\affil{I.Physikalisches Institut der Universit\"at zu K\"oln, Z\"ulpicher Str. 77, 50937 K\"oln, Germany}
\affil{Max-Plank-Institut f\"ur Radioastronomie, Auf dem H\"ugel 69, 53121 Bonn, Germany}
\author{Marzieh Parsa}
\affil{I.Physikalisches Institut der Universit\"at zu K\"oln, Z\"ulpicher Str. 77, 50937 K\"oln, Germany}

\begin{abstract}

We present the Keplerian orbit of S62 around the supermassive black hole SgrA* in the center of our Galaxy. We monitor this S-star cluster member over more than a full orbit around SgrA* using the Very Large Telescope with the near-infrared instruments SINFONI and NACO. For that, we are deriving positional information from deconvolved images. We apply the Lucy Richardson algorithm to the data-sets. The NACO observations cover data of 2002 to 2018, the SINFONI data a range between 2008 and 2012. S62 can be traced reliably in both data-sets. Additionally, we adapt one KECK data-point for 2019 that supports the re-identification of S62 after the peri-center passage of S2. With $t_{period}\,=\,9.9$ yr and a periapse velocity of approximately $10\%$ of the speed of light, S62 has the shortest known stable orbit around the supermassive black hole in the center of our Galaxy to date . From the analysis, we also derive the enclosed mass from a maximum likelihood method to be $4.15\,\pm\,0.6\,\times\,10^6 M_{\odot}$. 

\end{abstract}


\section{Introduction}
With the development of better instrumentation and observational techniques, the immediate environment of the supermassive black hole Sagittarius A* (SgrA*) can be investigated in detail. One of the fundamental quantities that determines the nature of the supermassive black hole SgrA* is its mass.
This quantity can be determined by using gaseous and stellar probes. An early mass determination was done by \cite{Wollman1977} who found an enclosed mass of $4\,\times\,10^{6}\,M_{\odot}$ through observations of the $12.8\,\mu m$ NeII line emission, origination from the mini-spiral located in the Galactic Center stellar cluster. Wollman's measurements used ionized gas as mass probes. Therefore, it could not be fully excluded that pressure gradients or magnetic fields could have influenced the derived quantity.
Also, it was unclear at the time, how much the central stellar cluster would contribute to the derived enclosed mass.
Therefore, radial velocities (\citealp{Krabbe1995}) and stellar proper motions of individual stars \citep[][]{Eckart1996a, Eckart1997, Ghez1998} allowed a much clearer insight into the amount of the compact mass and hence the gravitational potential associated with the supermassive black hole SgrA*.\\
The authors used the Viral theorem and the Jeans equation to derive the distance towards SgrA*. Due to this approach the enclosed mass could only be determined at a minimum distance from SgrA* typically given by the mean separation of the half dozend closest S-stars \citep[see][for the nomenclature]{Eckart1997}. The situation improved when single stars could be used via the detection of curvatures in their orbital tracks \citep[][]{Ghez2002, Eckart2002}. Complete orbits of the star S2 \citep[][]{Schoedel2002, Horrobin2004, Eisenhauer2003, Gillessen2009} then allowed to measure the enclosed mass down to the periapse distance of this star. Recently, the star S2 could be followed through its periapse passage using GRAVITY at the VLTI interferometer \citep[][]{Gravity2019}. With an orbital analysis of the innermost stars in the S-cluster \citep[the region around SgrA* with a diameter of 1".0, see also][]{Eckart1996a}, the mass estimate now settled around a value of 4.15 million solar masses with an uncertainty of about 0.2 million solar masses \citep[][]{Boehle2016, Parsa2017, Gravity2019}.
\\
The increased precision of the GRAVITY instrument \citep{gravity} allowed a mass determination of $4.148\,\pm\,0.014\,\times\,10^{6}\,M_{\odot}$ and the distance of $8175\,\pm\,13$ pc. The minimum distance between S2 and SgrA* during the periapse passage was
17 light hours (120 AU) at an orbital period of 16 years and an ellipticity of $0.88429\,\pm\,0.00006$. Searching for even closer stars resulted in finding the star S0-102 with an 11.2 year period and an ellipticity of $0.68\,\pm\,0.02$ with a correspondingly larger distance from SgrA* \citep{Meyer2012}.\\
Furthermore, mass probes were possible by investigating hot plasma blobs orbiting the super-massive black hole SgrA*. Here, \cite{Karssen2017} were able to measure the mass enclosed within 15 gravitational radii (R$_g$; $1\,\sim\,R_g\,\sim\,5\,\mu as\,=\,0.2\,\mu$ pc) by modeling the profiles of the brightest X-ray flares. Tracking hot-spots in the near-infrared using the K-band interferometric instrument GRAVTIY mounted at the VLTI \citep[located in Paranal/Chile, see also][]{gravity2018} gave a mass estimate at a distance of only 8 R$_g$.
It is unclear, how far these estimates derived from ionized hot gas blobs are influenced by viscosity or magnetic fields. Therefore, it would be more effective to use stars that are much closer than the periapse separation achieved by the star S2 to probe the gravitational potential of SgrA*.
\\
For this work, we will apply the iterative Lucy Richardson algorithm to our NACO and SINFONI data. With this high-pass filter, we de-blurr images and can separate and track stars with high angular resolution. With the results of this analysis technique, we derive the shortest known stable stellar orbit to S62 around SgrA* to date. This star has been identified earlier by \citep{Gillessen2009}. We fit a Keplerian orbit to the data and calculate the $\chi^2$ values in order to discuss the quality of the fit. For the orbital fit of S62 around the SMBH in the Galactic center, we use the fitting techniques presented in \cite{Parsa2017}.
In the following Sec. \ref{sec2}, we will explain the observation and the used instruments at the VLT. We introduce the applied analysing techniques and discuss the orbit fit. Section \ref{sec3} presents the results of the analysis and is followed by a discussion with a final conclusion in Sec. \ref{sec4}. In the Appendix, we list the used data, describe the analysing tools in detail, and present the re-identification of S62 during and after the peri-center passage of S2 in 2018.37 \citep{Gravity2019}.

\section{Observations $\&$ Analysis}
\label{sec2}
In this section, we will present the observations that are carried out with the Very Large Telescope in Paranal/Chile. We also give an overview of the data used (see also Appendix A) and introduce the analysing tools.

\subsection{NACO $\&$ SINFONI}

We are using the near-infrared instrument NAOS+CONICA (NACO) in imaging mode mounted at the VLT \citep[][]{Lenzen2003, Rousset2003} with the K-band filter. The used Adaptive Optics (AO) Laser guide star (LGS) IRS7 ($mag_K \,=\,7.7$) is located around $5".5$ north of SgrA*. The target is randomly dithered within a given area of $4".0$. Each exposure consists of 3 integrations of 10 seconds each. The reduction procedure for the data is described in \cite{Witzel2012}, \cite{Shahzamanian2016}, and \cite{Parsa2017}. We use the standard data reduction procedures including sky subtraction, bad-pixel and flat-field correction. Parts of the reduced data is also used in \cite{Witzel2012}, \cite{Eckart2013a}, \cite{Shahzamanian2016}, and \cite{Parsa2017}.

The near-infrared integral field spectrograph SINFONI \citep[][]{Eisenhauer2003a} is, like NACO, mounted at the VLT. The here used data is downloaded from the ESO archive\footnote{www.eso.org}. The data is observed in the H+K grating and the smallest plate scale of 0".025 with an exposure time of 600 seconds/single observation (see Tab. \ref{tab:data_sinfo1}). Since the wavefront sensor of SINFONI works only in the optical, the selection for the AO is limited to a star that is located 15".54 north and 8".85 east of SgrA*. Because the magnitude of the star is at the allowed limit ($\sim14\,$ mag), successful observations depend strongly on the weather conditions. To improve the efficiency of the observations, the AO loop could be opened during the night to do a re-acquisition of the guide star. This would improve the quality of the data by a factor of $10\%$ to $20\%$. This is done by measuring the PSF of S2 for stable seeing conditions.
The sky observations are done on a dark cloud located at 5'.36" north and 12'.45" west of SgrA*. The B2V star S2 is centered in the upper right quadrant to avoid a non-linear behavior of the detector (see the SINFONI user manual\footnote{www.eso.org}). The standard observational pattern is object (o) - sky (s) - object (o). Every other pattern besides the o-s-o setting influences the data because of the fast sky-variability in the infrared domain during the observations (\citealp[see][for a detailed discussion]{Davies2007, Peissker2019}) and is therefore excluded from the analysis. 
Standard G2V stars are observed for the telluric correction. We apply a flat field correction since some slits (usually slit 15 and 16) suffer from increased brightness features. Because the edges of some data cubes show errors that can not be corrected, we crop these regions by flagging the individual single data cubes. 
\\
After the corrections are applied, we select just single data cubes with a S2 PSF size of $<7.0$ pixel in both spatial directions. These selected single cubes are shifted in a $100\,\times\,100$ pixel array to a reference position that is defined from a previous created reference frame. From this, the final data-cubes are created from the combination of the corrected single exposures/cubes. 

\subsection{High-pass filtering}
\begin{figure*}[htbp!]
	\centering
	\includegraphics[width=1.\textwidth]{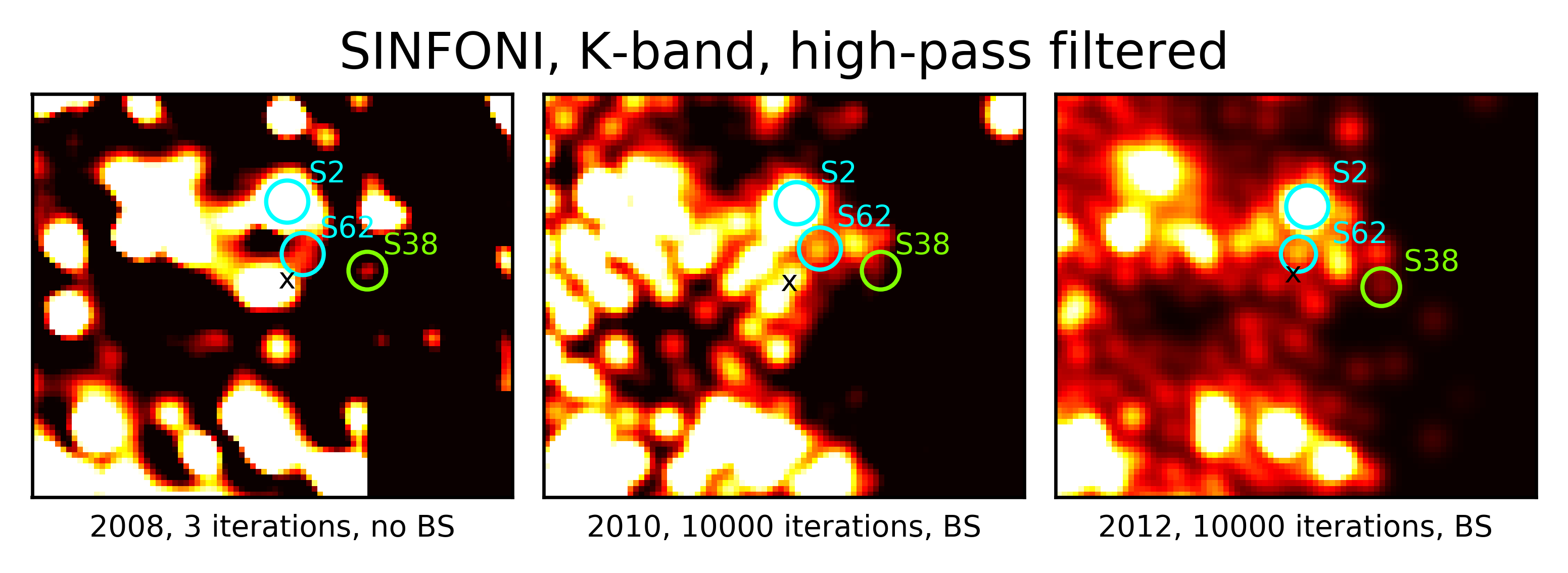}
	\caption{Detection of S62 in the SINFONI data-set. The black cross marks the position of SgrA*. S2, S38, and S62 are indicated by circles. North is up, East is to the left. The size of the FOV in all three images is 1".0 x 0".9. For display purposes, the lookup table is different from image to image. The images before deconvolution are shown in Appendix D.}
\label{fig:sinfo}
\end{figure*}
\begin{table*}[htbp!]
    \centering
    \begin{tabular}{cccccccc}
            \hline
            \hline
            Source & a [mpc] & e & i [$^\circ$] & $\omega$ [$^\circ$] & $\Omega$ [$^\circ$]&  $t_{closest}$ [years]& $t_{period}$ [years] \\
            \hline
            S62   & 3.588 $\pm$ 0.002 &  0.976 $\pm$ 0.002 & 72.76 $\pm$ 4.58 & 42.62 $\pm$ 0.4 & 122.61 $\pm$ 0.57 & 2003.33 $\pm$ 0.01 & 9.9 $\pm$ 0.02 \\    
            \hline
    \end{tabular}
    \caption{Orbital parameters for S62. The 1$\sigma$ uncertainty is based on the variation of R.A. and DEC. values by $\pm$6.5 mas.}
    \label{tab:orbital_para}
\end{table*}

The Lucy-Richardson alogrithm \citep{Richardson1972, Lucy1974} can be used to highlight image details. A high-pass filter is one solution to distinguish suppressed signals from emission that is caused by the background as well as the detector. Since the GC is a crowded region with a structured and variable background \citep{Sabha2012}, the deconvolution greatly supports the identification of a distribution of point-like stars. A detailed description is also presented in \cite{Peissker2019}. The technique, which derives from Bayes' theorem on conditional probabilities, conserves the constraints on spatial frequency distributions and, at each iteration, increases the likelihood of the resulting deconvolved image to represent the observed image. \cite{Ott1999} compared the flux density conserving properties of the Lucy-Richardson, the Clean, and the Wiener deconvolution algorithms for point-like stars in the Galactic Center environment. The robustness of the algorithm is demonstrated in Appendix B.
In Fig. \ref{fig:sinfo}, we present 3 deconvolved images. We extract K-band images from the collapsed SINFONI data-cube (see Fig. \ref{fig:sinfo_raw} in Appendix D) and apply, if necessary, a static background-subtraction (BS). This BS is adjusted to the data-quality, the background of the object, and the detector noise. Because of not avoidable superposition effects in the crowded GC regarding the PSF, we are using an artificial PSF (APSF). This APSF is created and modified with respect to the observed "natural" PSF of S2, the brightest source in the SINFONI FOV. In general, the SINFONI PSF is often rotated and shows an elongated shape with short and long axis values between 4-7 px depending on the data-quality. The quality of the deconvolved images depend on the match of the APSF to the real PSF and the correct choice of the background-subtraction. A large number of iterations is required to allow the algorithm to converge to a stable solution at all flux density levels. The effect can be seen in Fig. \ref{fig:sinfo}. The robustness of the S62 source detection in the framework of artificial source-planting and PSF subtraction is discussed in Appendices B and E.

\subsection{Orbital fit}

For the orbital fit of S62, stellar positions for each epoch of observation are measured using the NACO data and QFitsView (Thomas Ott, MPE Garching). The position of SgrA* in the NACO data is based on the well known orbit of S2 \citep[see][]{Gillessen2009S2, Parsa2017, Gillessen2017} in combination with the orbital elements. Based on the orbit and the offset-positions of S2, we can determine the position of SgrA* \citep{Parsa2017, Gravity2019} in order to create the reference frame. From this, we extract the offset position of S62 to SgrA*. This procedure can be also applied to the SINFONI data. The position of SgrA* is consistent with observed flares in the H+K-band and the SiO masers.

For the Keplerian fit, we are using the minimizing and iterative method L-BFGS-B \citep[see e.g.][]{saputro2017} for handling the bound constraints. This memory friendly algorithm is suitable for box constraints. We fit the semi-major axis, the eccentricity, the inclination, the periapsis, the longitude, and the time for the closest approach with respect to the starting time of the algorithm.

We obtained a starting value (initial guess) for the orbital elements by varying 
the R.A. and DEC. values of the measured positions by $\pm$6.5~mas,
averaging the results, and determined the 1$\sigma$ uncertainties.
Then we allow the elements to vary randomly within their 3$\sigma$ limits. 
We bootstrap the solution by using 50 representations of the
randomized elements calculating the resulting orbits and the deviation 
from the measured data. 
Following this approach, we obtained the best fitting orbit and the 
uncertainties from the uncertainty weighted distribution of the 
orbital elements.

From \cite{Gravity2019}, we use a mass for SgrA* of $4.15\,\times\,10^{6}\,M_{\odot}$. Because we are using stellar offset positions, the location of the SMBH is centered in the origin of the reference frame. 
Due to variations in the line of sight background of the S-stars and its relative position with respect to bright neighboring sources, the stellar positions measured from single epoch images do not necessarily show a Gaussian distribution. For each year, we are therefore using the median position\footnote{The median is less sensitive to outliers
and in case of a Gaussian distribution the median equals the mean.} of S62 whenever possible. With this approach, we minimize the effect of outlying data points and pay tribute to the variable background (see also simulations by Sabha et al. 2012 and the Appendices B and C in this work).

\section{Results}
\label{sec3}
In this section, we will show the results of our Keplerian fit and the enclosed mass of SgrA*.
\begin{figure*}[htp!]
	\centering
	\includegraphics[width=1.0\textwidth]{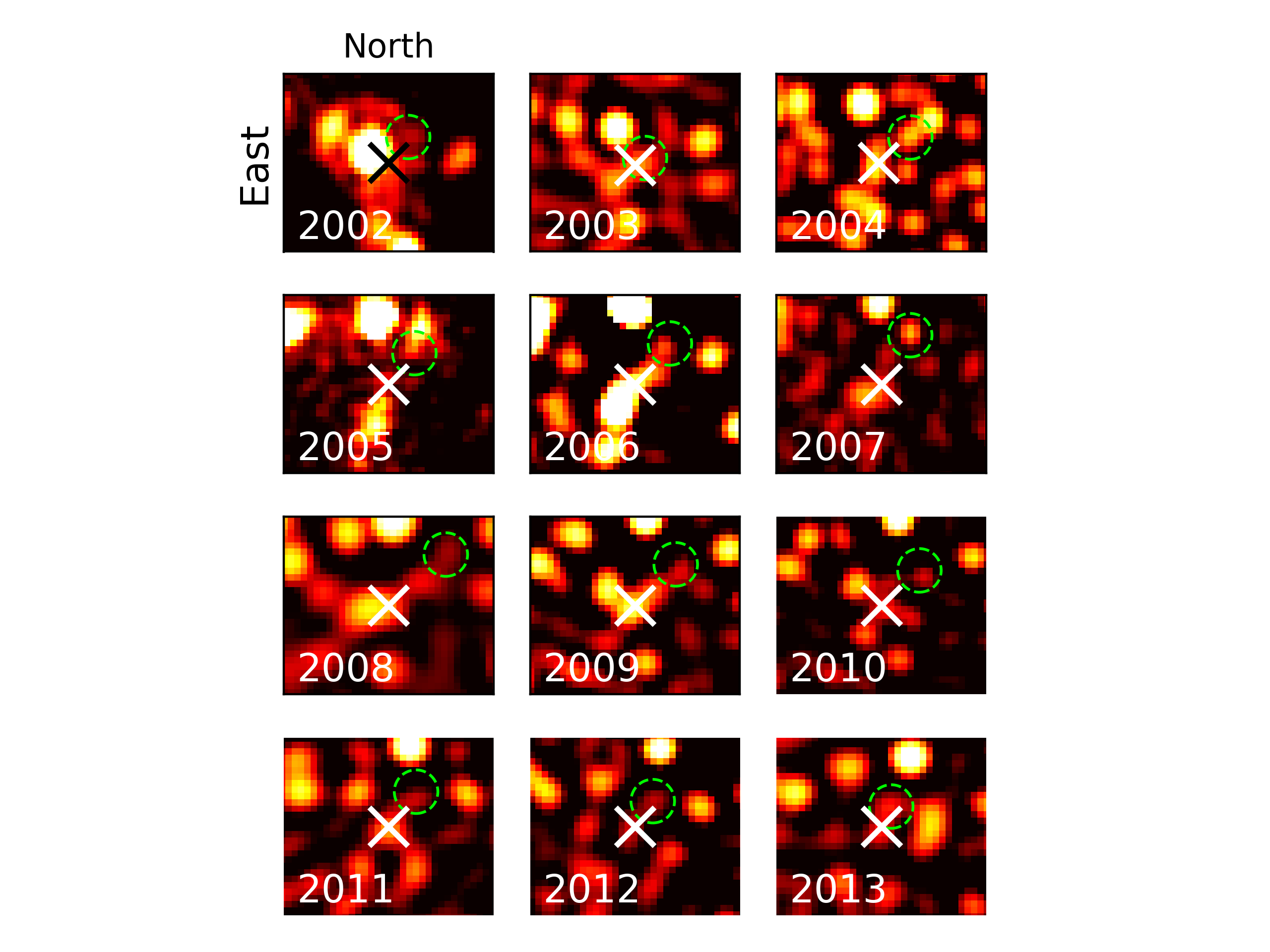}
	\caption{Selected overview of the S62 position in the NACO data around SgrA* between 2002 and 2013. The white cross indicates the position of SgrA*, S62 is located at the position of the lime colored dashed circle. The images are centered on SgrA*. The size of the FOV is 0".42 x 0".36. The angular resolution of the images is at the $60$ mas diffraction limit of the telescope in K-band. For the re-identification of S62 after the S2 passage through the field of view in 2014-216 see Appendix C.}
\label{finding_chart}
\end{figure*}
\begin{figure*}[htp!]
	\centering
	\includegraphics[width=0.9\textwidth]{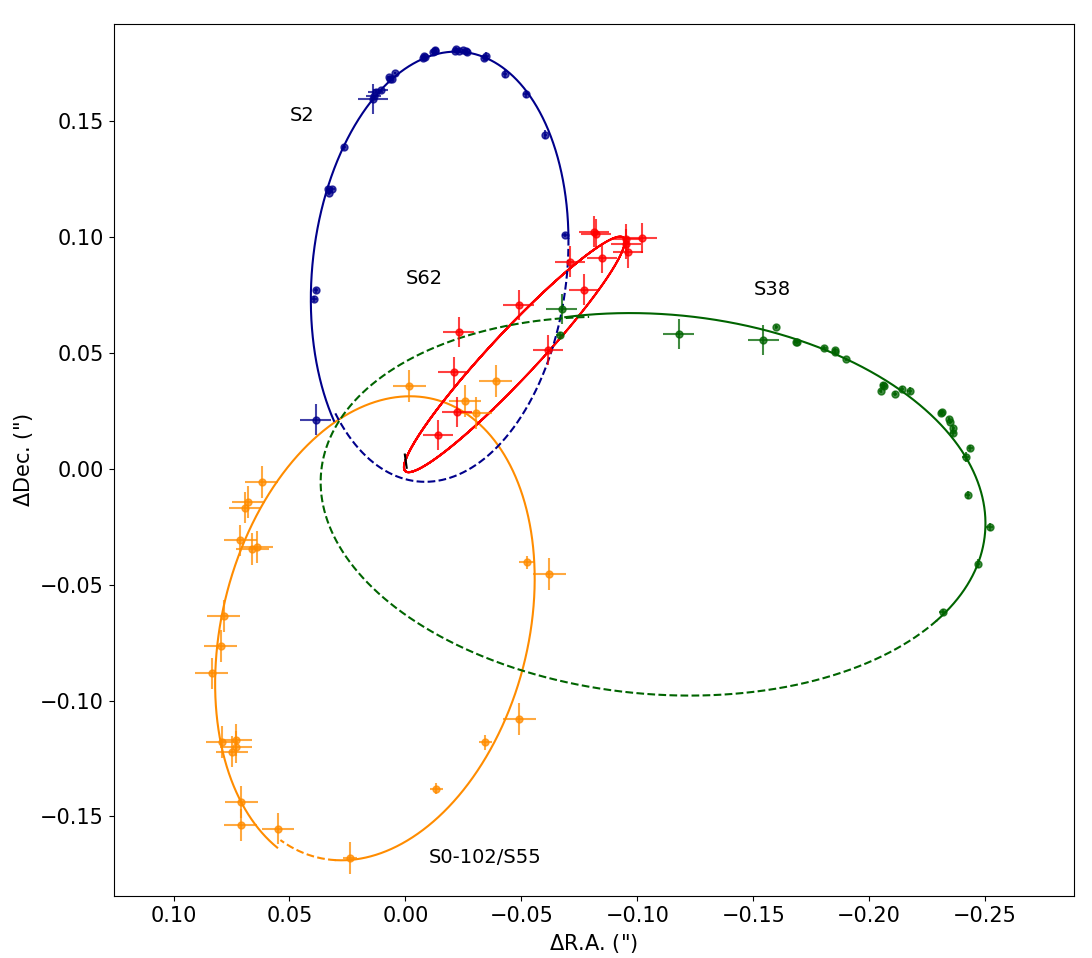}
	\caption{Orbit overview of S2, S38, S62, and S0-102. SgrA* is located at the origin of the coordinate system. See also Fig. 3 in \cite{Parsa2017}.}
\label{fig:all_stars}
\end{figure*}
From the NACO and SINFONI images, we derive the distance between SgrA* and S62 for every analysed data-set. They are based on deconvolved K- $\&$ H+K band images. The Keplerian-fit results in a 9.9 year orbit of S62 around SgrA* (Fig. \ref{finding_chart}) and is based on the NACO detection and the KECK data point adapted from \cite{Do2019} (see also Appendix C). The resulting orbital parameters can be found in Tab. \ref{tab:orbital_para}. To underline the robustness of the fit, we include the SINFONI data (red data-points) of 2008, 2010, and 2012.
\begin{figure}[htpb!]
	\centering
	\includegraphics[width=.44\textwidth]{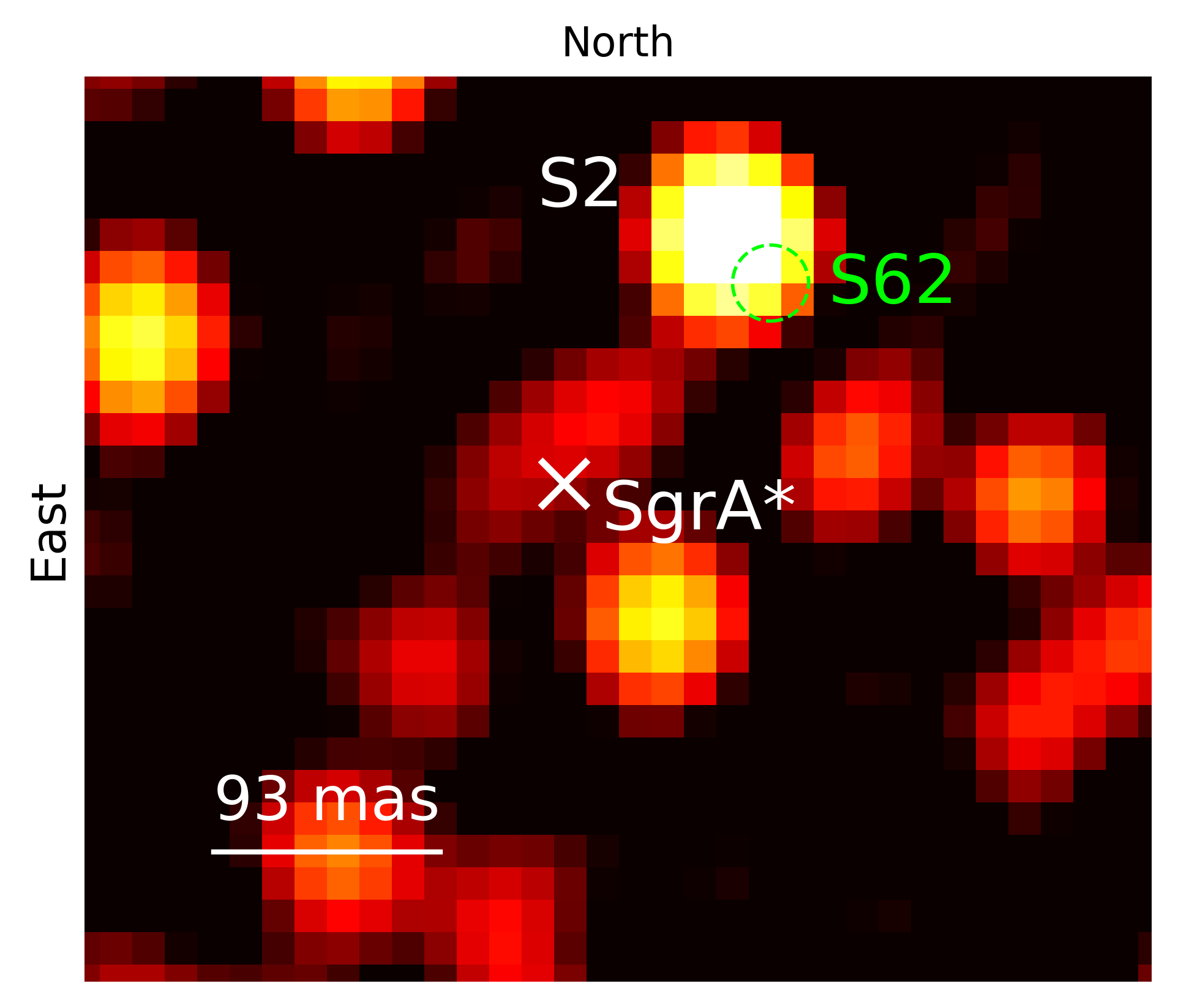}
	\caption{The GC in 2015. SgrA* is located at the white cross, S62 is expected to be at the location indicated by a lime dashed circle. S62 is too close to S2 and therefore not observable.}
\label{fig:2015_orbit}
\end{figure}
\begin{table*}[htbp!]
    \centering
    \small
    \tabcolsep=0.11cm
    \begin{tabular}{ccccccc}
    \hline
    \hline
    & \multicolumn{2}{c}{NACO} &\multicolumn{2}{c}{SINFONI} & \multicolumn{2}{c}{KECK}\\
    date & $\Delta$R.A. [mas]  & $\Delta$Dec. [mas] & $\Delta$R.A. [mas]  & $\Delta$Dec. [mas] & $\Delta$R.A. [mas]  & $\Delta$Dec. [mas] \\
    \hline
    2002.57 & -22.35  $\pm$ 5.64  &  24.37  $\pm$ 5.51 & -                 & -                  & - & - \\
    2003.44 & -14.04  $\pm$	4.77  &  14.56  $\pm$ 5.21 & -                 & -                  & - & - \\
    2004.51 & -61.32  $\pm$	4.42  &  51.38  $\pm$ 4.92 & -                 & -                  & - & - \\
    2005.42 & -77.09  $\pm$	5.04  &  77.41  $\pm$ 5.62 & -                 & -                  & - & - \\
    2006.72 & -102.05 $\pm$	3.99  &  99.71  $\pm$ 4.66 & -                 & -                  & - & - \\
    2007.25 & -96.00  $\pm$ 5.64  &  93.40  $\pm$ 5.51 & -                 & -                  & - & - \\
    2008.26 &      -              &          -         & -92.50  $\pm$ 5.5 &  95.00  $\pm$ 8.5  & - & - \\
    2008.36 & -95.09  $\pm$	4.77  &  97.11  $\pm$ 5.21 & -                 & -                  & - & - \\
    2009.50 & -95.03  $\pm$ 5.64  &  99.02  $\pm$ 5.51 & -                 & -                  & - & - \\
    2010.36 & -70.80  $\pm$	4.77  &  89.50  $\pm$ 5.21 & -88.12  $\pm$ 4.6 &  92.50 $\pm$ 7.2   & - & - \\
    2011.34 & -48.77  $\pm$	4.42  &  70.92  $\pm$ 4.92 & -                 & -                  & - & - \\
    2012.37 & -22.92  $\pm$ 5.64  &  59.11  $\pm$ 5.51 & -                 & -                  & - & - \\
    2012.49 &      -              &          -         & -27.50  $\pm$ 3.4 &  65.00 $\pm$ 5.45  & - & - \\
    2013.49 & -20.74  $\pm$	4.77  &  41.89  $\pm$ 5.21 & -                 & -                  & - & - \\
    2017.50 & -81.38  $\pm$	2.16  &  102.44 $\pm$ 3.23 & -                 & -                  & - & - \\
    2018.35 & -84.89  $\pm$	1.33  &  90.87  $\pm$ 1.20 & -                 & -                  & - & - \\
    2019.37 &     -     	  &        -        & -                        & -                  & -82.27 $\pm$ 5.00 & 101.26  $\pm$ 5.00 \\
    \hline
    \end{tabular}
    \caption{Stellar positions of S62 for our SINFONI and NACO data. The uncertainties for the NACO and SINFONI data are based on the Gaussian fit of the source itself and equals an average of about $\pm$0.5 pixels corresponding to
    6.25 mas (SINFONI) respectively 6.5 mas (NACO). For the data point in 2019, we also make use of one KECK data point presented in\citep[see Fig. 1,]{Do2019}. In consistency with our NACO and SINFONI data, we adopted a positional uncertainty of $\pm$5~mas in each direction.}
    \label{tab:s62-pos}
\end{table*}

\subsection{Orbit}
We derive a highly eccentric orbit with an eccentricity of $e\,=\,0.976\,\pm\,0.002$. We also find the closest point of S62 to SgrA* with $\sim\,2\, mas$. This corresponds to around 215 R$_S$.

Since the first observed periapse of S2 in 2002, the orbit of S62 can be observed and analyzed. With a K-band magnitude of around $\sim\,14$mag, S2 is the brightest member of the S-cluster. Stars, that have positions close to S2, are therefore blended. Because S62 is on a highly eccentric 9.9 year orbit, the observations after 2013 show an confusion with the S2 orbit (see Fig. \ref{fig:all_stars} and Fig. \ref{fig:2015_orbit}). It can be concluded, that S62 is only observable without blending after the periapse of S2 in a time-period for around 11 years. This is sufficient to cover one full orbit of the S-star S62.
\begin{figure*}[htbp!]
	\centering
	\includegraphics[width=.7\textwidth]{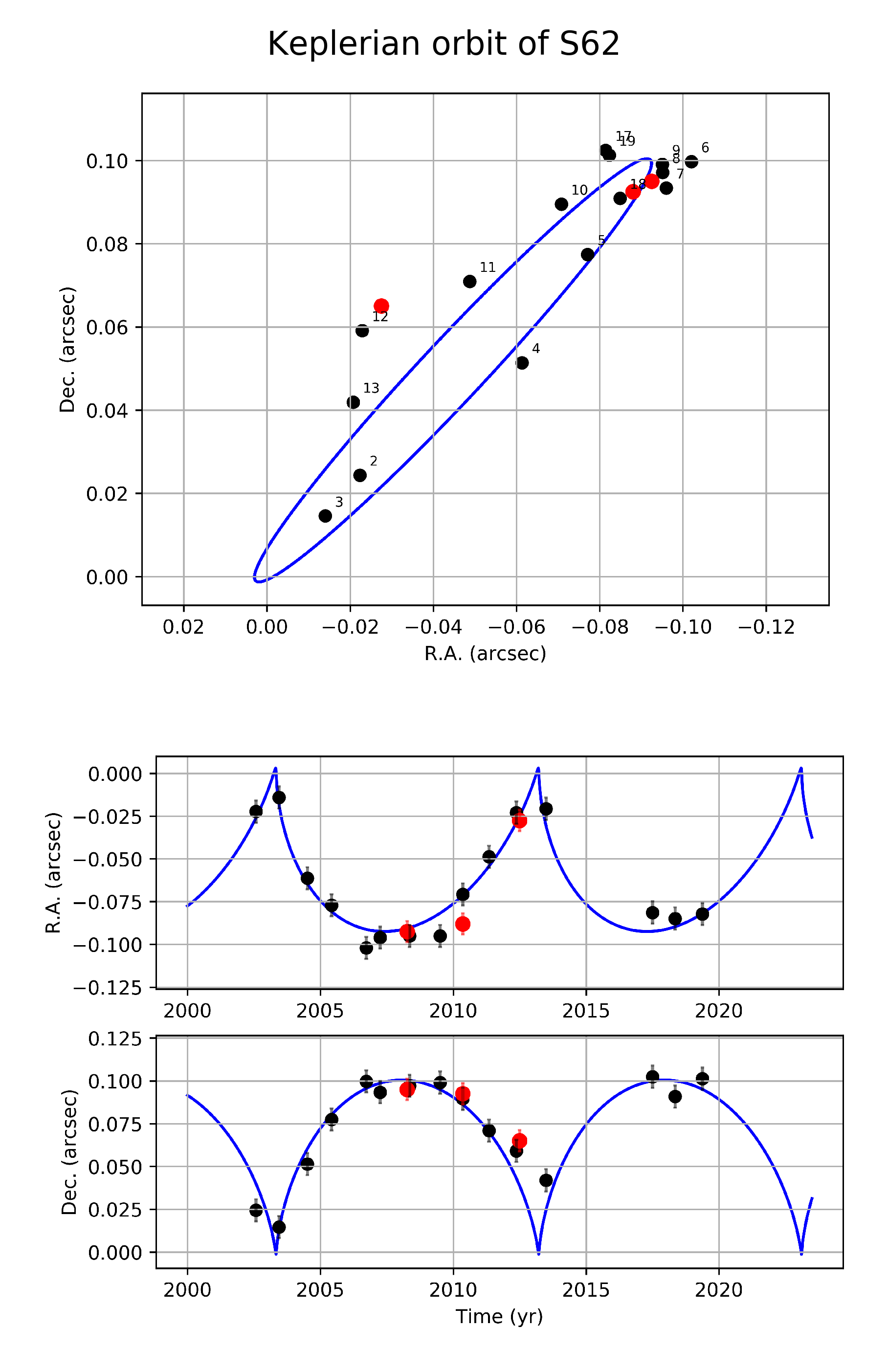}
	\caption{Keplerian orbit of S62. The data points are based on our NACO and SINFONI analysis with an error of $\pm\,6.5$ mas. The black numbers in the upper plot represent the related year of the data-points (+2000 yr). SINFONI data is represented by the red dots, NACO positions are plotted with black circles. The peri-center passage is determined to be in $2003.33\,\pm\,0.02$ with a orbital period time of 9.9 years. The next peri-center passage of S62 is expected in $2023.09\,\pm\,0.02$.}
\label{orbit}
\end{figure*}
\begin{figure}[htp!]
	\centering
	\includegraphics[width=.44\textwidth]{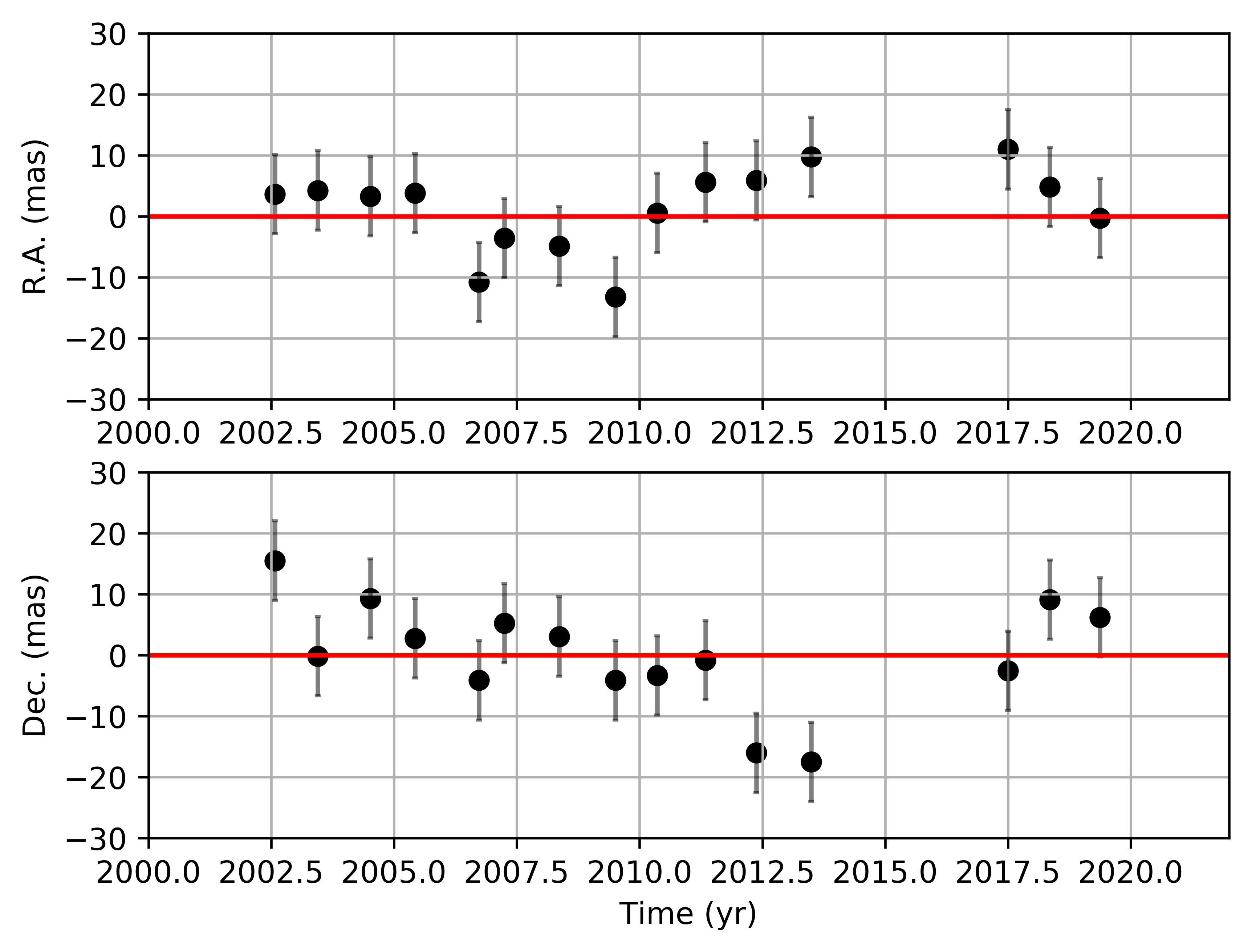}
	\caption{Residual plot of the fitted orbit data. The spatial pixel size is 13 mas, the error is adapted from the Keplerian orbit fit presented in Fig. \ref{orbit}.}
\label{fig:residuals}
\end{figure}
To highlight the robustness of the fit, Fig. \ref{fig:residuals} shows the residual plot of the orbital fit. The error bars (about $\pm\,0.5$ px $=\,\pm\,6.5$ mas) are adapted from the positional uncertainty (see uncertainties in Tab. \ref{tab:s62-pos}) of the orbit plot. The standard deviation for the R.A. plot is $3.75$ mas and for the DEC. plot $8.59$ mas and therefore in good agreement with the positional error of $\pm\,6.5$ mas.

\subsection{Minimized likelihood}
\label{sec:likeli}
\begin{figure}%[htp]
	\centering
	\includegraphics[width=.43\textwidth]{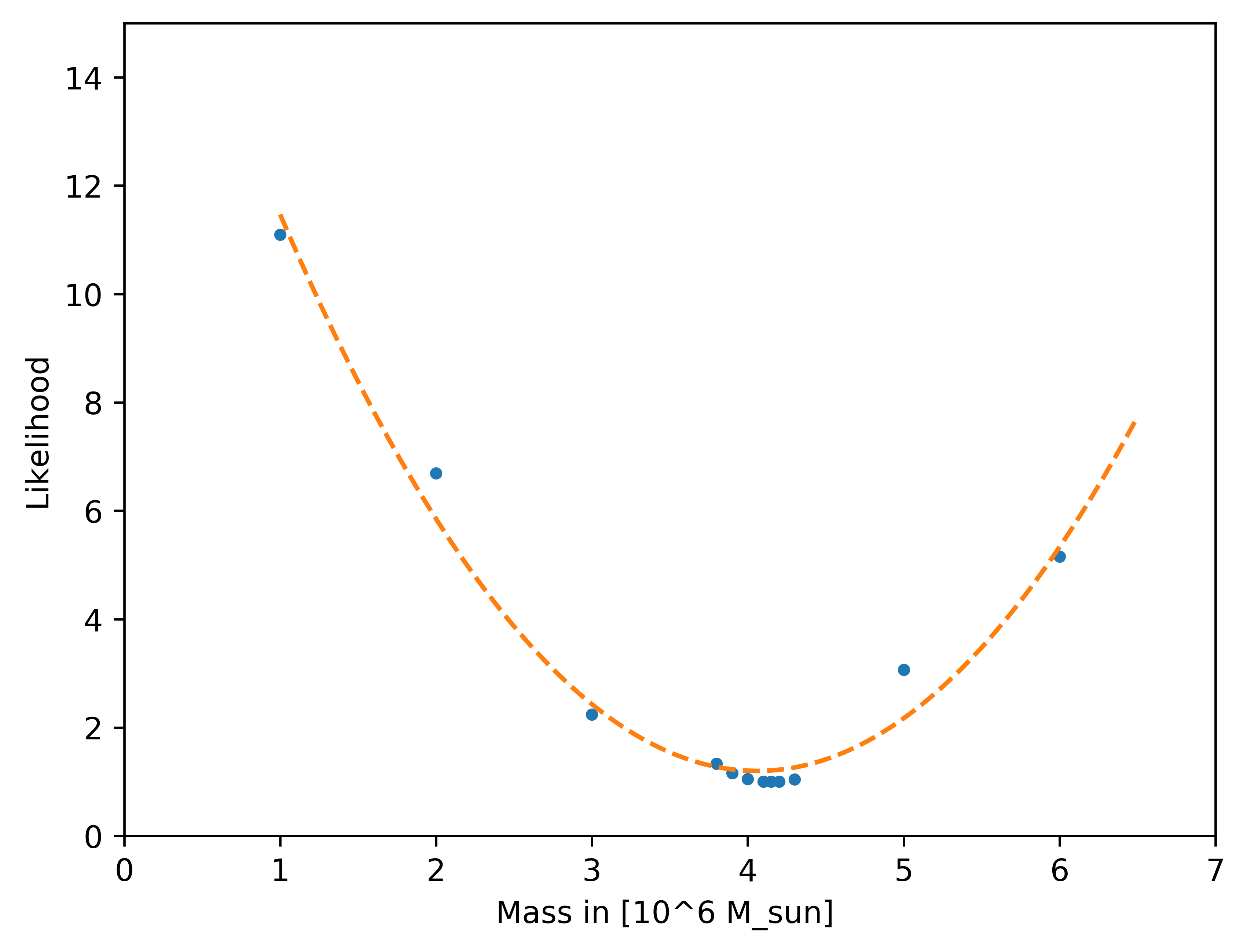}
	\caption{Likelihood as a function of mass. We normalized the resulting likelihood fit values to the minimum value of the mass variation analysis.}
\label{fig:likelihood}
\end{figure}

The minimize function for the six orbital elements as a function of the mass of SgrA* returns the residuals squared of the (initial guess) parameters.
The likelihood function can be interpreted as an indicator for the goodness of the fit \citep{Parsa2017} since it calculates the sum of the squared residuals. From the minimized and optimized orbital fit parameters, we are introducing a variation of the SMBH mass. Based on the analysis and the fit, the resulting plot of the likelihood function as a function of the mass of SgrA* is showing indeed a minimum at $(4.15\,\pm\,0.6)\,\times\,10^{6}\,M_{\odot}$ which is consistent with the mass derived by \cite{Gravity2019}. 
The  uncertainties of the mass is derived from the range of mass values for which the variation of the value of the $\chi^2$ value is below unity. Hence, we find for the central mass $(4.15\pm0.6)\times\,10^6\,M_{\odot}$ (see Fig. \ref{fig:likelihood}). This value is in good agreement with the mass derived by \cite{gravity2018, gravity2018b, Gravity2019}.

\subsection{Enclosed mass}

\begin{figure*}[htbp!]
	\centering
	\includegraphics[width=1.\textwidth]{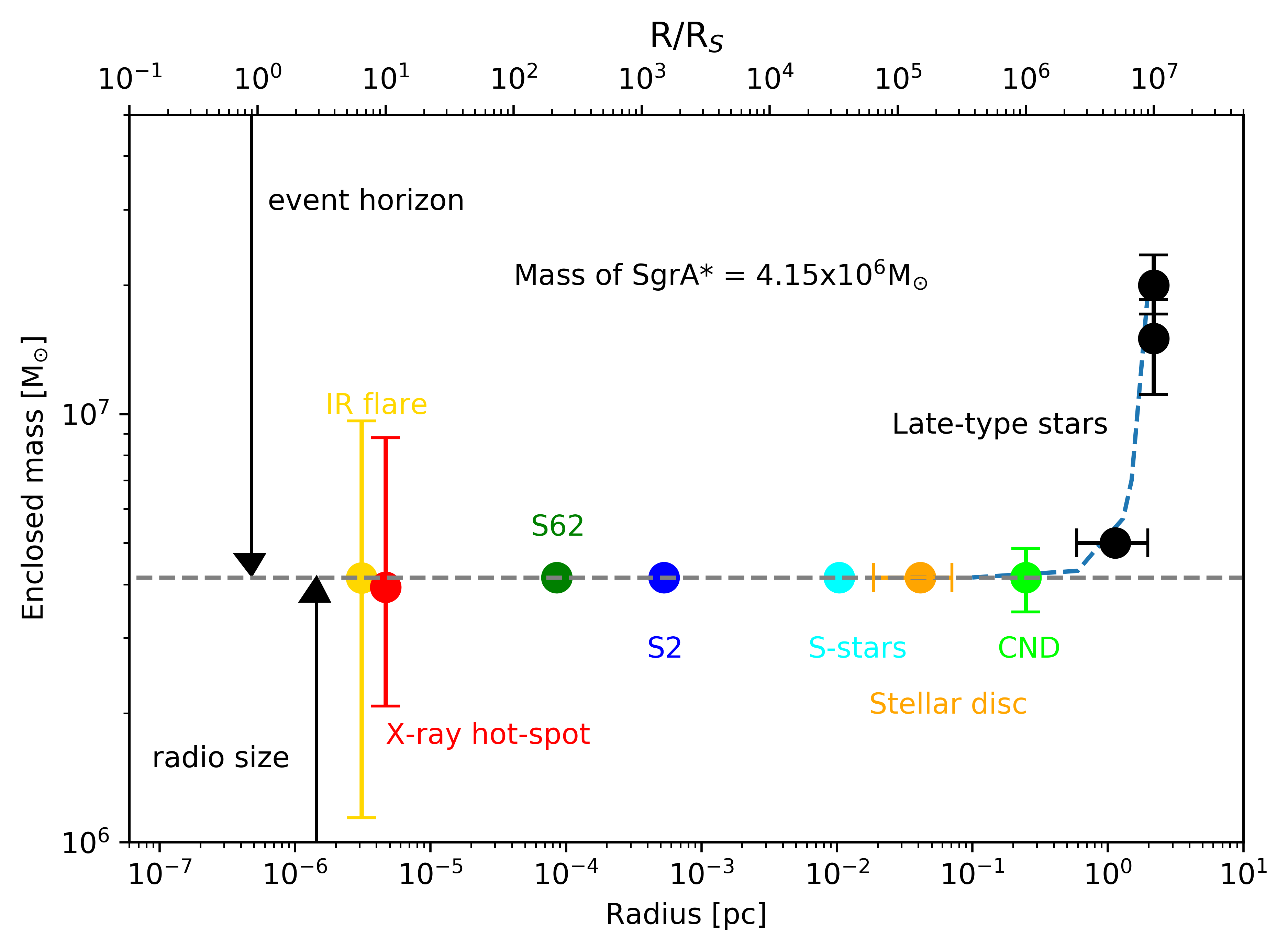}
	\caption{Enclosed mass of several objects. The first vertical dashed line marks the event horizon of SgrA*, the second one the measured radio size of the black hole. The uncertainties of the X-ray hot-spot is based on \cite{Karssen2017}. If no error bar is shown, the uncertainty of the mass determination is at or below the thickness of the symbol. The IR-flare error is at $4.14^{1.36}_{-1.14}\,\times\,10^6 M_{\odot}$ \citep{gravity2018b}. See also Fig. 5.1.1 in \cite{Genzel2010}.}
\label{fig:enclosed mass}
\end{figure*}

As described in Sec. \ref{sec:likeli}, we find a maximum for the minimize function that defines the enclosed mass. This value is based on a identification that marks the closest observed separation from a stellar source on a stable orbit to a supermassive black hole to date.
S62 is located significantly closer to the SMBH than to S2 \citep[see][]{Gillessen2009S2, Parsa2017, Gillessen2017} or S0-102 \citep{Meyer2012}. We can compare our result with other known objects in the GC (see Fig. \ref{fig:enclosed mass}). The hot-spot estimation as well as the error is based on the modeling of \cite{Karssen2017}. The authors use scale free orbiting hot-spot modeling that is based on the shape of observed flares. From this, they derive the mass of the central black hole associated with SgrA* after introducing the observed flare length in seconds. The authors find a value of $3.9^{+4.8}_{-1.8}\times\,10^6 M_{\odot}$ enclosed within 15 $R_g$, i.e. 7.5 Schwarzschild radii (3.0 $\mu$pc). \cite{Gravity2019} observed infrared hot-spots orbiting SgrA* at a separation of 6-10 $R_g$, i.e. 3-5 Schwarzschild radii (1.2-2 $\mu$pc). Based on the GRAVITY observations, the authors of \cite{Gravity2019} derive an enclosed mass of $4.15\,\pm\,0.01\,\times10^6 M_{\odot}$.
The values for the CND, S2, Stellar disk, and late-type stars in Fig. \ref{fig:enclosed mass} are adopted from \cite{Genzel2010}.

\section{Discussion $\&$ Conclusion}
\label{sec4}
This section summarizes the findings. We will discuss the results and give an brief outlook to upcoming observations. We also compare S62 to the hot-spot model.

\subsection{Properties of S62}

We compare the peak counts of S2 and S62 in 2012 with a one-pixel aperture. From this, we find a factor of 6 between both stars. With
\begin{equation}
    {mag}_{S62}\,=\,{mag}_{S2}\,-\,2.5*log(count-ratio)
\end{equation}
where the count-ratio is proportional to the flux-ratio of the two stars, we derive a K-band magnitude for S62 of around ${mag}_{S62}\,=\,16.1\,mag$. This result is in line with the other S-stars \citep[see][for more information]{Gillessen2009, Parsa2017, Gillessen2017, Cai2018} and lets us determine the mass of S62 with the mass-luminosity relation to about $2.2\,M_{\odot}$ if we assume a main-sequence star and a index of $a\,=\,3.5$.  
\newline
While \cite{Jalali2014} have shown that stars of a few solar masses can  be formed in the immediate vicinity of an SMBH, it is unlikely that S62 has formed on such a highly elliptical orbit that brings it so close to the SMBH.
Stellar scattering provides a likely scenario for placing a star in
a closely bound orbit around SgrA*.
According to the Hills scenario, a binary with a total mass $m_b$ and
a semimajor axis $a_b$ is tidally disrupted whenever it comes close to 
a super massive black hole with mass $M_{SMBH}$ within a distance of the order of the tidal disruption radius

\begin{equation}
r_t[AU] \approx 10 \frac{a_b}{0.1~AU} {\left(\frac{M_{SMBH}}{m_b/4 M_{\odot}}\right)}^{1/3}~~~.
\end{equation}

For the Galactic Center, this implies a 20 $M_{\odot}$ binary originating from
within the 1" diameter S-star cluster. This binary must have had a 
semimajor axis of about 1~AU in order to get disrupted.
This disruption results in a hyper-velocity star and a star, that is
even harder bound to the SMBH after the disruption event compared to its initial state \citep{Hills1988, Rasskazov2019, Sari2019}.

For stars very close to the SMBH, collisions become more likely than scattering events. In this domain, the orbital velocities exceed the typical escape velocity from the surface of a star (around 600 $km s^{-1}$ for a solar mass star). 
Such a collision may result in a merger or the disruption of the star. This, however, depends on the ratios between their masses, their encounter velocities,
surface escape velocities, and the impact parameter \citep{Benz1987, Trac2007, Gaburov2010, Alexander2017}.
While a collision is indeed a very likely fate of S62 in the near future, the object itself is unlikely to be a product of a disruptive collision. In that case, the stellar material would have been stretched out over a large section of the orbit very quickly. The event as such would result in an, at least temporary, luminous and extended trail.
S62, however, appears to be of similar brightness compared to other S-star cluster members and is very compact, i.e., not extended on scales resolvable by the angular resolution of the telescope.

However, rather than a collision, a future tidal disruption is also very likely.
\cite{Fragione2019} give the tidal disruption radius of a SMBH-MS binary with respective masses $m_{SMBH}$ and $m_{MS}$ as

\begin{equation}
R_T = R_* \left( \frac{m_{SMBH}}{m_{MS}} \right)^{1/3}~~~.
\end{equation}

The radius of a star more massive than $> 1.66 \, M_{\odot}$ can be obtained from

\begin{equation}
R_* [R_{\odot}] = 1.33 (m_{MS}/M_{\odot})^{0.555}\, ,
\end{equation}

see \cite{Demircan1991}.

Assuming a mass of 2.2 $M_{\odot}$ for S62 (see above), we find $R_* = 2.06 R_{\odot}$ and a tidal disruption radius of $251 R_{\odot}$ or $1.18 AU$. This can be compared to the periapse separation $r_p$. Using the orbital elements in Tab. \ref{tab:orbital_para} we find:
\begin{equation}
   r_p = a(1-e) \approx 16~AU~~.
\end{equation}
Hence, the periapse distance of S62 is about 3 times the hot-spot distance as determined by \cite{Karssen2017} and just about 15 times the tidal disruption radius. Certainly, tidal interactions will already be important for the evolution of S62. The derived and observed properties of S62 indicate that it is indeed an S-star member on a tight orbit around the supermassive black hole SgrA*.

\subsection{Gravitational periapse shift}

We determine a minimum distance of S62 to SgrA* that is comparable to about 30 times the distance of as determined with the hot-spot model presented by \cite{Karssen2017}. Figure \ref{fig:enclosed mass} shows, that S62 is an excellent candidate to probably show relativistic effects presented in \cite{Parsa2017} and \cite{gravity2018, Gravity2019}. Since the orbital time scale is measured to be 9.9 years, the next peri-center passage will be around March 2023. During that passage, the star will have a velocity of about $10\%$ the speed of light.

All experiments indicate, that the mass associated with SgrA* is very compact and is most likely presented in the form of a super-massive black hole \citep{Eckart2017}.
In this case, the relativistic {\em prograde} peri-center advance per
revolution is given by (see \cite{Weinberg1972}, Eq. (8.6.11))

\begin{equation}\label{adv}
\Delta\varphi=\frac{6\pi G}{c^2}\, \frac{M}{a(1-e^2)},
\end{equation}

\noindent
with $a$ being the semi-major axis and $e$ the eccentricity of the orbit,
respectively.
\\
Using the values from the orbital elements for S62, we find a periapse advance of $9.9^{\circ}$. Given the value of the argument of the peri-center $\omega$, a significant fraction of the periapse advance should be measurable on the sky. Since a full orbit tracking should be out of phase with S2 probably every
second orbit of S62, the star can be observed and used to derive the periapse advance.

The expected shifts in position due to the relativistic peri-astron shift are currently well in the positional uncertainties of our measurements. However, future interferometric observations with GRAVITY will considerably improve the result. 

\section*{Acknowledgements}
We are greatful to Basel Ali (I. Physikalisches Institut, Cologne) for support with the source identification and the related orbit calculations. We also thank the members of the NACO/SINFONI and ESO Paranal/Chile team.
This work was supported in part by the
Deutsche Forschungsgemeinschaft (DFG) via the Cologne
Bonn Graduate School (BCGS), the Max Planck Society
through the International Max Planck Research School
(IMPRS) for Astronomy and Astrophysics, as well as special
funds through the University of Cologne and
SFB 956: Conditions and Impact of Star Formation. We thank the Collaborative Research Centre 956, sub-project [A02], funded by the Deutsche Forschungsgemeinschaft (DFG) – project ID 184018867. 

\bibliography{bib}

\appendix
\label{sec:app}
\section*{\Large Appendix A: Data}

\begin{table*}[h!]
\centering
\begin{tabular}{ccccc}
\hline
\hline
\multicolumn{5}{c}{NACO}\\
\hline
Date (UT) & Observation ID & \multicolumn{1}{p{1.5cm}}{\centering number \\ of exposures }   & \multicolumn{1}{p{1.5cm}}{\centering Total \\ exposure time(s) } & $\lambda$ \\
\hline
2002-07-31 & 60.A-9026(A)  & 61  & 915     & K \\
2003-06-13 & 713-0078(A)   & 253 & 276.64  & K \\
2004-07-06 & 073.B-0775(A) & 344 & 308.04  & K \\
2004-07-08 & 073.B-0775(A) & 285 & 255.82  & K \\
2005-07-25 & 271.B-5019(A) & 330 & 343.76  & K \\
2005-07-27 & 075.B-0093(C) & 158 & 291.09  & K \\
2005-07-29 & 075.B-0093(C) & 101 & 151.74  & K \\
2005-07-30 & 075.B-0093(C) & 187 & 254.07  & K \\
2005-07-30 & 075.B-0093(C) & 266 & 468.50  & K \\
2005-08-02 & 075.B-0093(C) & 80  & 155.77  & K \\
2006-08-02 & 077.B-0014(D) & 48  & 55.36   & K \\
2006-09-23 & 077.B-0014(F) & 48  & 55.15   & K \\
2006-09-24 & 077.B-0014(F) & 53  & 65.10   & K \\
2006-10-03 & 077.B-0014(F) & 48  & 53.84   & K \\
2006-10-20 & 078.B-0136(A) & 47  & 42.79   & K \\
2007-03-04 & 078.B-0136(B) & 48  & 39.86   & K \\
2007-03-20 & 078.B-0136(B) & 96  & 76.19   & K \\
2007-04-04 & 179.B-0261(A) & 63  & 49.87   & K \\ 
2007-05-15 & 079.B-0018(A) & 116 & 181.88  & K \\ 
2008-02-23 & 179.B-0261(L) & 72  & 86.11   & K \\ 
2008-03-13 & 179.B-0261(L) & 96  & 71.49   & K \\ 
2008-04-08 & 179.B-0261(M) & 96  & 71.98   & K \\ 
2009-04-21 & 178.B-0261(W) & 96  & 74.19   & K \\ 
2009-05-03 & 183.B-0100(G) & 144 & 121.73  & K \\ 
2009-05-16 & 183.B-0100(G) & 78  & 82.80   & K \\ 
2009-07-03 & 183.B-0100(D) & 80  & 63.71   & K \\ 
2009-07-04 & 183.B-0100(D) & 80  & 69.72   & K \\ 
2009-07-05 & 183.B-0100(D) & 139 & 110.40  & K \\ 
2009-07-05 & 183.B-0100(D) & 224 & 144.77  & K \\ 
2009-07-06 & 183.B-0100(D) & 56  & 53.81   & K \\ 
2009-07-06 & 183.B-0100(D) & 104 & 72.55   & K \\ 
2009-08-10 & 183.B-0100(I) & 62  & 48.11   & K \\ 
2009-08-12 & 183.B-0100(I) & 101 & 77.32   & K \\
\hline  
\end{tabular}
\caption{First part of the used K-band NACO data. For every epoch, the number of exposures used for the final mosaics, the total exposure time,and the Project ID is listed. Note that NACO was decommissioned between 2013 and 2015.}
\label{tab:naco_data1}
\end{table*}
\begin{table*}[hp!]
\centering
\begin{tabular}{ccccc}
\hline
\hline
\multicolumn{5}{c}{NACO}\\
\hline
Date (UT) & Observation ID & \multicolumn{1}{p{1.5cm}}{\centering number \\ of exposures }   & \multicolumn{1}{p{1.5cm}}{\centering Total \\ exposure time(s) } & $\lambda$ \\
\hline
2010-03-29 & 183.B-0100(L) & 96  & 74.13   & K \\ 
2010-05-09 & 183.B-0100(T) & 12  & 16.63   & K \\ 
2010-05-09 & 183.B-0100(T) & 24  & 42.13   & K \\ 
2010-06-12 & 183.B-0100(T) & 24  & 47.45   & K \\ 
2010-06-16 & 183.B-0100(U) & 48  & 97.78   & K \\
2011-05-27 & 087.B-0017(A) & 305 & 4575    & K \\
2012-05-17 & 089.B-0145(A) & 169 & 2525    & K \\
2013-06-28 & 091.B-0183(A) & 112 & 1680    & K \\
2017-06-16 & 598.B-0043(L) & 36  & 144     & K \\
2018-04-24 & 101.B-0052(B) & 120 & 1200    & K \\
\hline  
\end{tabular}
\caption{Second part of the used K-band NACO data.}
\label{tab:naco_data2}
\end{table*}

\begin{table*}[htbp!]
 	\centering
 	\begin{tabular}{cccccccc}
  	\hline
  	\hline
    \multicolumn{8}{c}{SINFONI}\\
    \hline
 	\\	Date & Observation ID & Start time & End time & \multicolumn{3}{c}{Amount and quality of the data} & Exp. Time \\  \cline{5-7} & & & & Total & Medium & High &  \\
 	(YYYY:MM:DD) &  & (UT) & (UT) &  &  &  & (s)  \\ \hline 
	
 	2008.04.06 & 081.B-0568(A) & 05:25:26 & 08:50:00 & 16 &   0  &  15   &    600  \\
 	2008.04.07 & 081.B-0568(A) & 08:33:58 & 09:41:05 &  4 &   0  &   4 &    600  \\
  	2010.05.10 & 183.B-0100(O) & 06:03:00 & 09:35:20 & 3 &   0  &  3   &    600  \\
 	2010.05.11 & 183.B-0100(O) & 03:58:08 & 07:35:12 &  5 &   0  &   5 &    600  \\
 	2010.05.12 & 183.B-0100(O) & 09:41:41 & 09:57:17 & 13 &   0  &  13  &    600  \\
 	2012.03.18 & 288.B-5040(A) & 08:55:49 & 09:17:01 &  2 &   0  &   2 &    600  \\
 	2012.05.05 & 087.B-0117(J) & 08:09:14 & 08:41:33 & 3 &   0  &  3  &    600  \\
 	2012.05.20 & 087.B-0117(J) & 08:13:44 & 08:23:44 & 1 &   0  &  1  &    600  \\
 	2012.06.30 & 288.B-5040(A) & 01:40:19 & 06:54:41 & 12 &   0  &  10  &    600  \\
 	2012.07.01 & 288.B-5040(A) & 03:11:53 & 05:13:45 & 4 &   0  &  4  &    600  \\
 	2012.07.08 & 288.B-5040(A)/089.B-0162(I) & 00:47:39 & 05:38:16 & 13 &   3  &  8  &    600  \\
 	2012.09.08 & 087.B-0117(J) & 00:01:36 & 00:23:33 & 2 &   1  &  1  &    600  \\
 	2012.09.14 & 087.B-0117(J) & 01:21:30 & 01:43:27 & 2 &   0  &  2  &    600  \\
 	\hline	\\
 	\end{tabular}	
 	\caption{SINFONI data of 2008, 2010, and 2012. For an overview, the total amount of data in the related years is listed. To ensure the best S/N ratio for our combined final data-cubes, we are just using single data-cubes with high quality. Dates are listed in UT.}
 	\label{tab:data_sinfo1}
 	\end{table*}
 	
\newpage
\section*{\Large Appendix B: Source crowding and noise}

Since the Galactic center is a very crowded region, we need to investigate in how far
our source identifications are compromised by the crowding effects.
In the following, we investigate the detection of sources in a crowded field, the influence of excess pixel noise on the deconvolution, and the probability of finding an serendipitous orbit.

\begin{figure*}[h!]
	\centering
	\includegraphics[width=1.\textwidth]{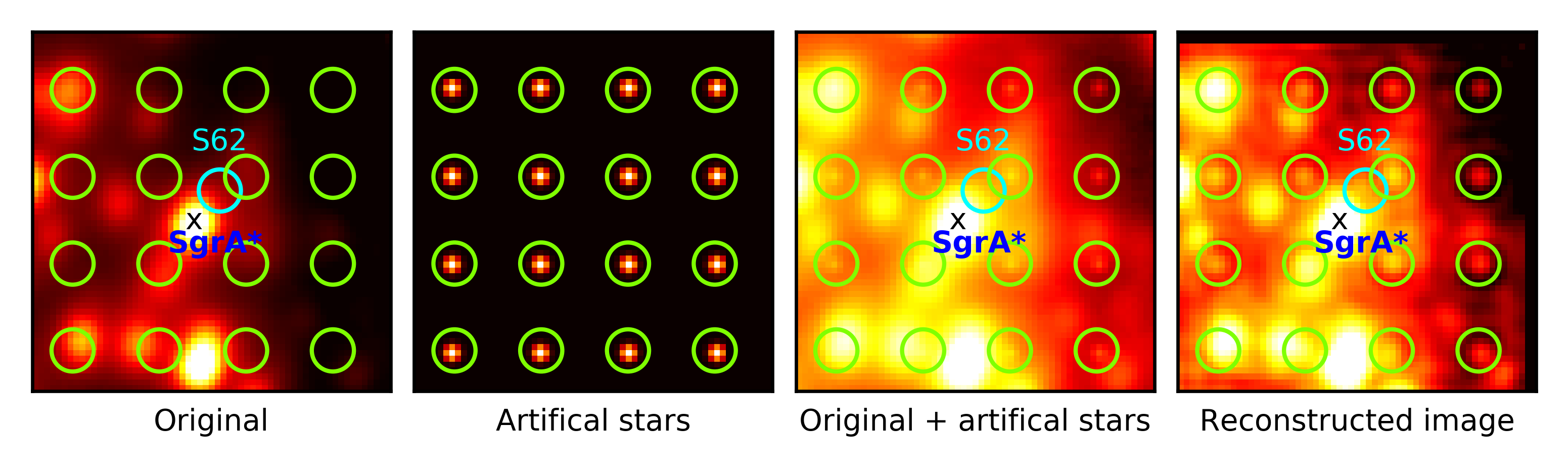}
	\caption{Example from the test on one particular field:
a) original image;
b) array of artificial stars; counting from
top left (1) to bottom right (16);
c) Original image with artificial stars planted;
d) Result after deconvolution and reconvolving to the final angular resolution close to the diffraction limit. 
Zone I cases: The artificial sources at the
field-position 1 and 13 could not be separated from the natural star close to that position.
Zone II cases: At the field-position 7 and 14, the planted artifical star can be
identified but is close to a natural star.
Zone III cases: In all other field-positions, the planted artifical stars fell at or beyond half a PSF width from a natural star and could clearly be separated.}
\label{fig:artifical_star}
\end{figure*}

\noindent
{\it Source detection in a crowded field:}
In order to test the liability of our source detection, we conducted source planting experiments. We created a 4$\times$4 array of artificial sources with the same noise
and flux properties as S62. We added this to individual images 
before deconvolution. After deconvolution, we tested if the artificially 
placed sources could be detected.
We randomly positioned the 4$\times$4 array on several images and repeated the
process.
The result of the investigation in the image can be divided into 3 different 
zones (see Fig. \ref{fig:artifical_star}):
\\
{\it Zone I:} Artificially planted sources, that are within a radius of 0.3 times the half-power width of detected (real) S-stars, could not be separated from the objects originally presented in the image. In this case, the sources within Zone I had a flux corresponding to the one from the planted source plus the flux of the original source at that position after the deconvolution.
{\it Zone II:} For a small region with a distance of 0.3-0.6 times the half-power width from an S-star, we could retrieve about 60$\%$ of the planted sources with mostly compromised positions and fluxes. Those position measurements, that may have been influenced by neighboring sources in Zone II, are suppressed as extreme values by choosing the median in order to combine the different stellar positions per epoch.
{\it Zone III:} For the entire region with separations of at least 0.6 times the halfpower width of detected sources, we could always separate the planted sources from the sources presented in the image. Since the angular velocity of S62 in the crowded lower half part of its orbit is well above 30 mas/yr, the chances of finding it in Zone I for the duration of an entire year are very small. Combined with the fact, that we discarded source detections with fluxes significantly larger than the S62 flux, our source detections are all located exclusively Zone III. However, depending on the time variable (since everything is moving), the local line of sight background and the distribution of brighter sources close to the line of sight (and corresponding gradients in the local background), the positional uncertainty may reach
$\pm$10 mas (i.e., about a sixth of a beam) for the median of several position estimates. This uncertainty may be higher for individual single epoch images (see simulations by \citealt{Sabha2012} and \citealt{EckartAA2013}). The authors derive the chance for a false positive detection of a few percent for sources, that are detected for 3 consecutive years. We can conclude from this analysis, that the uncertainty for detecting a false positive detection is significantly smaller than 1$\%$ since we detect S62 for 11 consecutive years.
\\
\\
{\it Influence of noise:}
All images had an exposure time that is sufficient in order to detect the source. The S-cluster was always positioned in regions 
of the detector array with extremely good cosmetics. There were no regions of defect pixels, that are close in size to the PSF. The likelihood of individually healthy pixels forming a simultaneous
positive fluctuation is proportional to $\sigma^N$, where $N$ is the number
of pixels characteristic for the size of a PSF. This is very unlikely to happen.
For 1$\sigma$ and N=10, the likelihood is already only $\times$10$^{-5}$.

The effect of single pixels with a statistically high count-rate should also be discussed:
The influence of single pixel excursions is strongly suppressed by the 
Lucy algorithm. The reason is, that the ratio of images to correct the
$(i-1)^{th}$ iteration in order to obtain the new $i^{th}$ iteration is convolved with 
the PSF (see equation (15) in \citep{Lucy1974}). Therefore, the effect of a single pixel is smeared out over the entire PSF. We did experiments with 5$\sigma$-8$\sigma$ excursions in the count rate and did not find any significant effect in the resulting deconvolved images.
Only if the flux in a single affected pixel approaches a good portion of
the flux contained in the faintest detected sources, they may become
a source of confusion. In our images, we did not find an excessive count-rate
excursion of single pixels that would have lead to such an effect.
\\
\\
{\it Probability of finding an serendipitous orbit:}
One may find indications for an orbital motion if sources of a suitable 
magnitude serendipitously occur close to the investigated orbital positions.
There are about 50 sources within the central arcsecond that are bright 
enough to derive and trace their orbit. Only a fraction of them has a
flux density compatible with S62. If we require the source 
to be identified within a single PSF, the likelihood of finding a source
serendipitously at a suitable position is about $K=50/(1''/0.060'')^2$=0.18.
To meet these condition independently for 12 times, the
likelihood is $K^{12}=1.2 \times 10^{-9}$, which is too small to be considered.

\section*{\Large Appendix C: Re-identification of S62}
Here, the primary goal is to show that, despite the complex field, S62 could be re-identified while S2 is approaching its periapse position. S62 is showing very little proper motion because of its apoapse position, while all other sources in the crowded field move.
\begin{figure*}[htbp!]
	\centering
	\includegraphics[width=1.\textwidth]{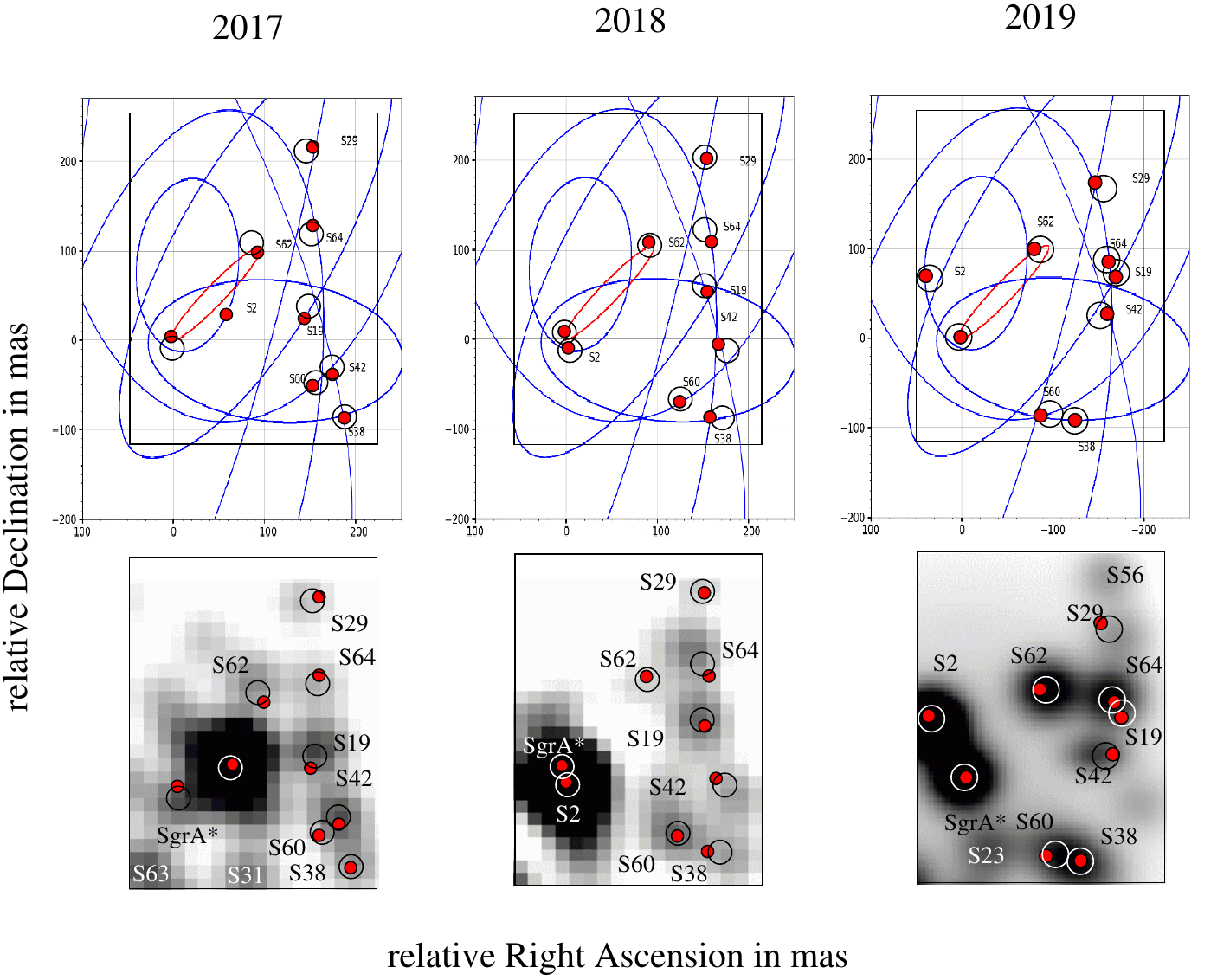}
	\caption{Re-identification of S62 after S2 passed through the field of view in 2014-2016. We clearly detect the star S62 in 2017, 2018, and 2019. {\it Top:} Predictions from orbital calculations based on data from 2002 till 2019 (see text for details).
{\it Bottom:} Single epoch images at the diffraction limit. The description of symbols is given in the text. The image scale in the top pannels are given in milliarcseconds. The rectangle in the top pannels outlines the FOV circumference of the lower pannels.}
\label{fig:reident}
\end{figure*}
Shortly before pericenter passage of S2 in 2018.37, the star passed through the S62 orbit as S62 was approaching its apocenter position. In 2014-2016, it was basically impossible to measure S62 due to the presence of the K=14 bright star S2. However, S62 could be re-identified in the crowded region between 2017 and 2019. Due to its high eccentricity of around 0.97, the S62 orbit is precisely determined and the position of the apocenter and the time for the apocenter passage are well known. For the sources close to S62, their trajectories could be calculated by performing orbital fits to the NACO data covering the years 2002 - 2018. We also adapted the data presented in \cite{Do2019}.
\begin{table}[htb]
\caption{Orbital elements of the indicated stars of Fig. \ref{fig:reident}.}
\begin{center}
\begin{tabular}{rrrrrrrrrrrr}\hline \hline
Star  &a(mpc)&	e  &	i($^o$)&$\omega$($^o$)&	$\Omega$($^o$)&	t$_{closest}$(yr)\\ \hline 
S2    &~5.04 $\pm$ 0.01   &0.884 $\pm$  0,002 &136.88 $\pm$ ~0.40  &~71.33 $\pm$  0.75  &234.51 $\pm$~1.03  &2002.32 $\pm$ 0.02\\
S19   &11.52 $\pm$ 1.98   &0.606 $\pm$  0,073 &~69.67 $\pm$ ~2.96  &139.00 $\pm$  5.96  &335.64 $\pm$~2.58  &2004.48 $\pm$ 0.01\\
S29   &28.69 $\pm$ 2.55   &0.476 $\pm$  0,095 &101.64 $\pm$ ~2.01  &350.70 $\pm$ 13.26  &170.00 $\pm$~2.07  &2046.98 $\pm$ 4.89\\
S38   &~5.63 $\pm$ 0.21   &0.804 $\pm$  0,050 &159.86 $\pm$ 15.01  &~15.70 $\pm$  9.65  &~98.43 $\pm$~8.31  &2003.33 $\pm$ 0.34\\
S42   &38.60 $\pm$ 2.75   &0.649 $\pm$  0,041 &~65.43 $\pm$ ~0.91  &~39.89 $\pm$  2.87  &206.32 $\pm$~2.24  &2012.29 $\pm$ 1.39\\
S60   &20.37 $\pm$ 3.22   &0.833 $\pm$  0,087 &132.43 $\pm$ ~6.42  &~50.31 $\pm$ 19.69  &206.40 $\pm$24.07  &2021.50 $\pm$ 4.99\\
S62   &~3.59 $\pm$ 0.01   &0.976 $\pm$  0,002 &~72.76 $\pm$ ~4.68  &~42.62 $\pm$  0.40  &122.61 $\pm$~0.57  &2003.33 $\pm$ 0.01\\
S64   &15.90 $\pm$ 2.71   &0.354 $\pm$  0,126 &114.21 $\pm$ ~1.80  &155.11 $\pm$ 31.35  &167.11 $\pm$~8.75  &2005.56 $\pm$ 5.27\\
\hline \hline
\end{tabular}
\end{center}
\label{tab:orbit_elements_additional_stars}
\end{table}
In Fig. \ref{fig:reident} we show the source arrangement in the years 2017, 2018, 2019 close to the S62 apocenter position. For 2017 and 2018, we used a single best NACO data set (see Tab. \ref{tab:naco_data2}) that allowed high angular resolution imaging in the region around SgrA*. For 2019, we show a Gaussian model representing the source distribution published by \cite{Do2019} in their Fig. 2. The comparison of the images with the resulting orbital calculations allow us to re-identify S62. In Fig. \ref{fig:reident}, we show relevant S-stars in the region just west of SgrA*: S19, S29, S38, S42, S60, and S64. We also indicate S23, S31, S56, and S63 in some years. 

In 2018, S62 seems to be fainter. It should be noted, that high resolution observations result in high sensitivity images. Sources in a crowded field are not constantly bright as a function of time (i.e. from year to year). This is due to the varying AO performance, the signal to noise, and most of all due to the variable background. As shown by \cite{Sabha2012}, significant variations on time scales of 1-3 years can be expected for fainter sources in the GC.

In 2017, the Lucy image reconstruction is difficult since S2 is very close to S62 and the neighboring sources to the west. Also, the region towards S64 appears to be confused. The red circles indicate the predicted positions of the sources. It has a width of $\pm$6.5 mas corresponding to the nominal uncertainty we reached for S62 (see Tab. \ref{tab:s62-pos}) based on several position measurements per epoch. The comparison between the predicted positions and the single epoch images in Fig.10 is hampered by 1) the uncertainties in the orbital elements and 2) by the scatter in the single epoch results depending on the line of sight background and the immediate vicinity of the sources. The black circles have a width of $\pm$13 mas corresponding to $\pm$1 pixel (i.e., $\pm$ one fifth of a diffraction limited beam) for the NACO camera in K-band (see also discussion in Appendix B). They are centered between the expected position and the actual peak position obtained for the single epoch image representation of the field.

\section*{\Large Appendix D: SINFONI K-band images of the GC}

\begin{figure*}[htbp!]
	\centering
	\includegraphics[width=1.\textwidth]{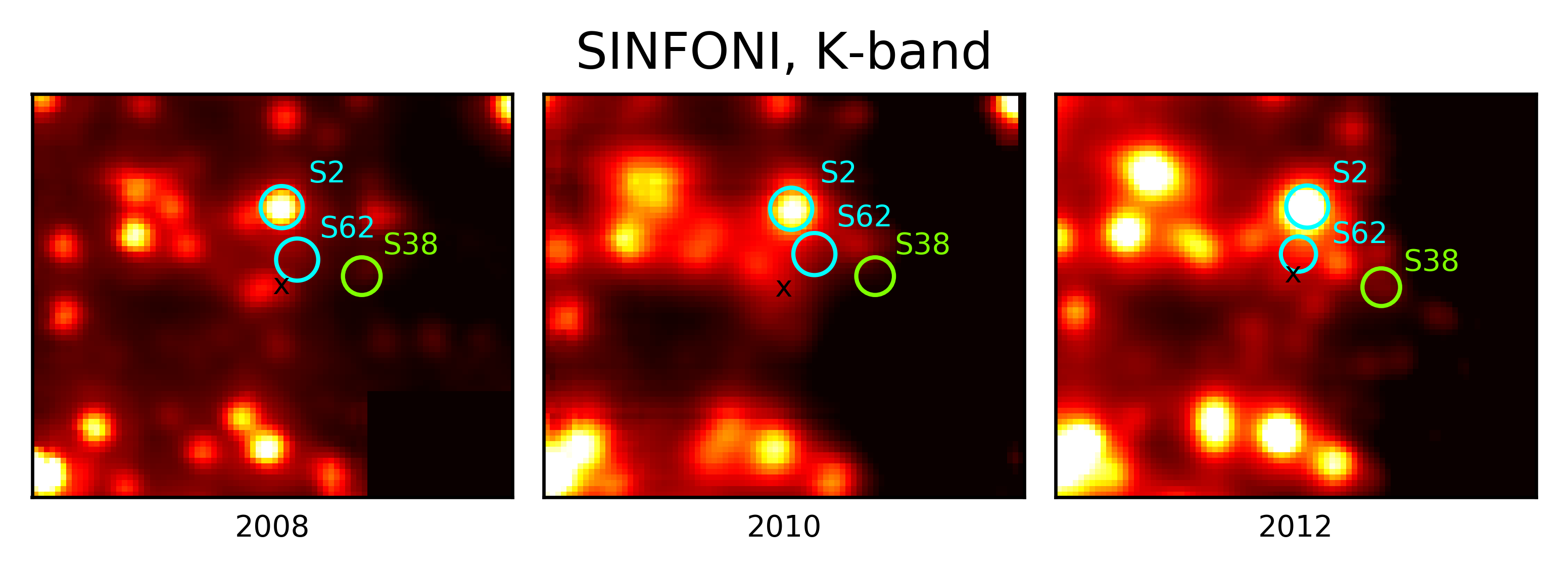}
	\caption{K-band images of the GC extracted from the collapsed SINFONI data-cubes. No Lucy-Richardson algorithm or spatial frequency filters have been applied to these images (see also caption of Fig. \ref{fig:sinfo}).}
\label{fig:sinfo_raw}
\end{figure*}

Here, we present the K$_S$-band images (Fig. \ref{fig:sinfo_raw}) that have been extracted from the SINFONI data-cubes. The on-source integration for 2008 is 210 minutes, for 2010 it is 640 minutes, and for 2012 we have a total of 610 minutes.

We shifted the marked S2 star to a fixed position. SgrA* is marked with a black {\it x}. 

\section*{\Large Appendix E: PSF subtraction}

\begin{figure*}[htbp!]
	\centering
	\includegraphics[width=1.\textwidth]{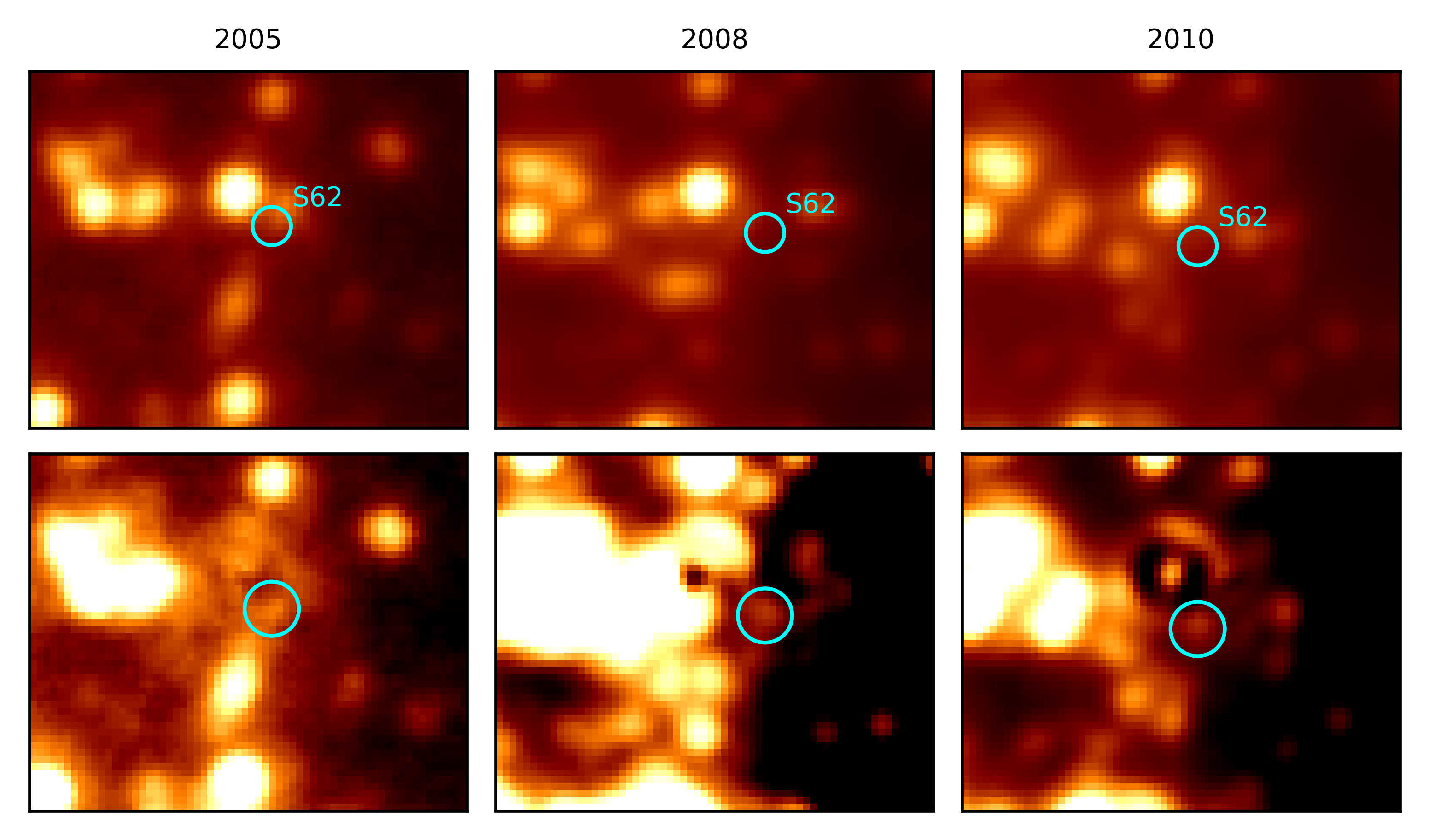}
	\caption{Galactic center in 2005, 2008, and 2010. North is up, East is directed to the left. The upper three NACO K-band images have been taken with the S13 camera (13 mas/spatial pixel scale) and show the S-cluster without any subtracted star. S62 is indicated by a cyan-colored circle. The lower three images show the PSF subtracted results in the corresponding years. The circle is at the same position and shows the source S62 at the same position as in the corresponding deconvolved images in Fig. \ref{finding_chart}.}
\label{fig:psf_sub}
\end{figure*}

Here, we demonstrate that the successful detection of the star S62 is
not an artifact of the deconvolution process. Furthermore, the source can be detected by subtracting bright sources in its vicinity.
In Fig. \ref{fig:psf_sub}, we show the results of the PSF subtraction. As shown, we subtracted several PSFs scaled to the brightness of surrounding stars in order to highlight the presence of S62. We use images from 2008, 2010, and 2012 as examples.
The NACO K-band images with all stars are aligned in the first row. The second row shows the detection of S62 after the subtraction of the nearby stars. The bright star close to the center is S2. Additional stars are labeled in Figures \ref{fig:sinfo}, \ref{fig:all_stars}, \ref{fig:2015_orbit}, and \ref{fig:sinfo_raw}.

\end{document}